\documentclass[fleqn,usenatbib]{mnras}

\usepackage[T1]{fontenc}
\usepackage{ae,aecompl}

\usepackage{graphicx}	
\usepackage{amssymb}	
\usepackage{longtable}

\usepackage[utf8]{inputenc}
\usepackage{pdflscape}
\usepackage{lscape}
\usepackage{booktabs}
\usepackage{csquotes}
\usepackage{color, colortbl}
\usepackage{amstext}
\usepackage{wasysym}
\usepackage{multirow}
\usepackage{natbib}
\usepackage{ulem}
\usepackage{subfigure}

\def\farcs{\hbox{$.\!\!^{\prime\prime}$}}

\usepackage{t1enc}
\usepackage{times}

\usepackage{graphicx}
\usepackage{footnote}
\usepackage{pifont}
\usepackage{lscape}
\usepackage{booktabs}

\usepackage{threeparttablex}
\definecolor{LightGray}{gray}{0.9}
\definecolor{LightGray1}{gray}{0.8}
\definecolor{pad}{rgb}{0.06,0.15,0.76}

\def\farcs{\hbox{$.\!\!^{\prime\prime}$}}

\hypersetup{draft} 

\title[First stellar perturbers of comet motion.]{First stars that could significantly perturb comet motion are finally found.}

\author[R. Wysocza\'{n}ska, P. A. Dybczy\'{n}ski, M. Kr\'{o}likowska]{Rita Wysocza\'{n}ska$^{1}$\thanks{E-mail: rita.wysoczanska@amu.edu.pl}, Piotr A. Dybczy\'{n}ski$^{1}$\thanks{E-mail: dybol@amu.edu.pl}, Ma{\l}gorzata Kr\'{o}likowska$^{2}$\thanks{E-mail: mkr@cbk.waw.pl}
\\
$^{1}$Astronomical Observatory Institute, Faculty of Physics, A.~Mickiewicz University, S{\l}oneczna 36, 60-286 Pozna\'{n}, Poland\\
$^{2}$Space Research Centre of the Polish Academy of Sciences, Bartycka 18A, 00-716 Warsaw, Poland
}

\date{Accepted XXX. Received YYY; in original form ZZZ}

\pubyear{2019}

\begin{document}
\label{firstpage}
\pagerange{\pageref{firstpage}--\pageref{lastpage}}
\maketitle

\begin{abstract}

Since 1950 when Oort published his paper on the structure of the cloud of comets it is believed that stars passing near this  hypothetical cometary reservoir play an important role in the  dynamical evolution of long period comets and injecting them into the observability region of the Solar System.  The aim of this paper is to discuss two cases in which the data obtained from observations were used and stellar perturbations  (of different intensity, strong case of C/2002~A3 LINEAR and weaker case of C/2013~F3 PANSTARRS) on cometary motion were detected. Using the best available data from the Gaia DR2 catalogue  and some other sources we searched for close stellar passages near the Sun. Our study took into account that some of the stars are parts of multiple systems. Over 600 stars or systems that approached or will approach the Sun closer than 4.0\,pc were found. Having the list of perturbers completed we studied their influence on a sample of 277 Oort spike comets that were observed since 1901 and discovered that two comets might have their orbits fundamentally changed due to a close stellar encounter. Our results show how much different the dynamical evolution of comets would have looked when their motion was considered only in the Galactic potential. Uncertainties both in stellar and cometary data were carefully taken into account. Our analysis indicates that the occurrence of stellar perturbations on cometary motions is very rare and the uncertainties of these effects are hard to estimate.    

\end{abstract}


\section{Introduction}\label{sec:intr}
It was proposed by \cite{oort:1950} that passing stars are responsible for making observable some long period comets (LPCs) residing in the postulated cometary reservoir. Since that time a continuing effort was given by many authors to find a star which perturbed an observed comet. While there exists a widely accepted opinion that close stellar passages near the Sun do occur, up to now no one could point to a particular star that perturbed past cometary motion in a significant manner. 

The search for close stellar encounters is restricted to stars with known right ascension $\alpha$, declination $\delta$, radial velocity $v_{r}$, parallax $\pi$, and proper motions $\mu_{\alpha^{*}}$,  $\mu_{\delta}$. For decades we were limited by a small number of measured stellar parallaxes and radial velocities. The HIPPARCOS mission \citep{hipparcos,vanleeuwen:2011} was a great advance in this respect, providing 120\,000 stellar parallaxes, but still a small fraction of these stars have their radial velocity measured, see for example the XHIP compilation catalogue \citep{anderson_francis:2011}. This was later improved by large observational programmes like RAVE \citep{RAVE-4:2013,RAVE-5:2017}. The situation reverses again with the Gaia mission \citep{Gaia_mission:2016}. While the Gaia DR2 catalogue \citep{Gaia-DR2:2018} provides us with five-parameter astrometry for over one billion stars, only a limited subset has radial velocities measured and a large number of stars that are bright and/or close to the Sun were omitted at this stage of the mission.

Using this growing amount of data several identifications of close stellar encounters with the Solar System were performed by many authors. Some of the most up-to-date studies on this topic were conducted by \cite{garcia-sanchez:2001,dyb-hab3:2006,jimenez_et_al:2011,dyb-kroli:2015,dyb-berski:2015}. The most recent published paper is by \cite{Bailer-Jones:2018}. Close stellar encounters are appealing to us because they allow us to investigate future and, more interestingly, past dynamics of the observed LPCs by a numerical integration of their motion in a three body problem (the Sun -- comet -- star) under the additional influence of the Galactic potential. To assess the influence of a star on a comet's motion as well as on the Sun and mutual star -- star interactions, we need also stellar masses which are not easy to find in the literature and often must be estimated in a crude manner. This issue will be addressed later in this paper.   

In Section \ref{sec:New_stars} we describe the procedure of collecting an up to date list of potential stellar perturbers while Section \ref{Increasing_LPCs_sample} describes a sample of LPCs for which we searched for stellar perturbations. We found the first examples of strong star -- comet interactions, which are described in detail in Sections \ref{sect:two_examples} -- \ref{sec:2012f3}. New perspectives and conclusions are in Sections \ref{sect:new_persp} -- \ref{sec:conclusions}.

\section{New stars thanks to Gaia mission}\label{sec:New_stars}

In April 2018 the second data release (DR2) \citep{Gaia-DR2:2018} from the Gaia mission \citep{Gaia_mission:2016} has been made available to the public. It allowed us to prepare a new and exhaustive, as never before, list of stars that once in the future or past could be found in the close vicinity of the Sun.  

Our attempt differs from a recent  paper by \cite{Bailer-Jones:2018} in several aspects:
\begin{itemize}
    \item First of all, we included stars not only from Gaia DR2 but also from other sources. All the stars previously pointed out in the above mentioned papers have been checked against the new Gaia astrometry and then the list has been extended with hundreds of new stars.  
    \item Second, we augmented Gaia DR2 radial velocities with values from other sources using the SIMBAD database \citep{SIMBAD-article:2000}. 
    \item Third, we took into account that some stars suspected to make a close stellar encounter with the Solar System are actually parts of binary or multiple systems and therefore it is necessary to calculate their center of mass kinematic parameters and use them instead of using these stars separately. 
    \item In addition, we used a more recent Galaxy potential model by \cite{irrgang_et_al:2013} which is sufficient for our purpose because we are interested in stars in a relatively close vicinity of the Sun, say a few hundreds parsecs. 
\end{itemize}

 We started with a list of all stars suspected to make a close encounter with the Sun, as it was pointed out in many previous papers \citep{garcia-sanchez:2001,dyb-hab3:2006,jimenez_et_al:2011,dyb-kroli:2015,dyb-berski:2015,bailer-jones:2015,Bailer-Jones:2018}.  Next the list was extended with new potential stellar perturbers selected using a linear motion approximation \citep[see for example][]{bailer-jones:2015,berski-dyb:2016}  from among the stars included in the Gaia DR2 catalogue with their $v_r$ from Gaia DR2 or from different source when available or more accurate. All candidates were then checked if they were single stars or members of a multiple system. In the latter case we collected data for other members and calculated the center of mass positions and velocities. 
 
Where it was possible we took five-parameter astrometry directly from Gaia DR2, adding $v_r$ either from Gaia DR2 itself or from other sources with a great help of the SIMBAD database \citep{SIMBAD-article:2000}. All numerical integrations were performed in the galactocentric frame. We obtained the position and velocity of the Sun and each star in that frame by using formulas and parameters gathered in \cite{berski-dyb:2016}.

For obtaining more reliable parameters of the stellar passages we numerically integrated the motion of each star or system in a two body problem (the Sun -- star)  in  the Galactic potential using the RA15 integrator  \citep{everhart-ra15:1985}. To be able to take into account the Sun -- star gravity we needed a mass of each star or system. A stellar mass was also necessary later in our main task -- examining the influence of these stars on a comet's motion. 

Masses were obtained from different sources. Over 400~stars from our list have their mass estimates presented in \cite{Bailer-Jones:2018}. For M dwarfs, where masses were hard to obtain from other sources, we used formulas from \cite{benedict2016solar}. For main sequence dwarfs with known effective temperature we estimated their mass using formulas included in \cite{eker2018interrelated} in conjunction with formulas from the Gaia documentation \citep{Gaia-DR2:2018} which allowed us to calculate the luminosity in different bands. Where possible we have checked the consistency between the luminosity and the effective temperature.  The TESS catalogue  \citep{stassun2018tess,muirhead2018catalog} was also used as a source of star masses. Additionally, for most of the stars it was possible to estimate their masses thanks to tables created by  \cite{Mamajek2013}. For most of the stars we have obtained several mass estimates, so we decided to take a 'reasonable mean' of those values after exclusion of the most extreme ones and those which were most flawed. When our list was filled with masses we could start further calculations. The integration was performed backwards or forwards for 50\,Myr  or alternatively until  the star or system heliocentric distance exceeded 3000\,pc. 

\begin{figure}
    \centering
    \includegraphics[width=\columnwidth]{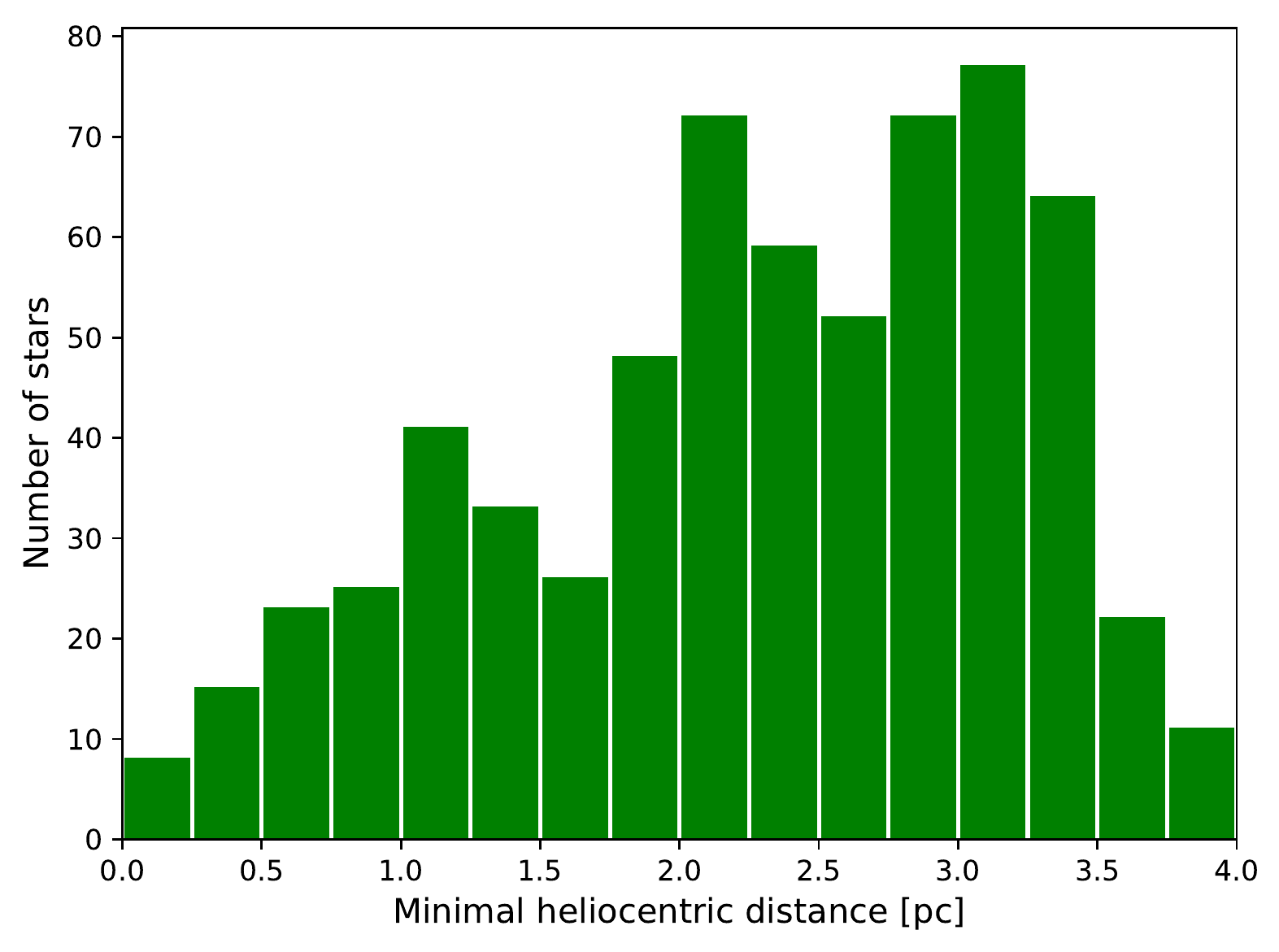}
    \caption{The distribution of minimal heliocentric distance for 647~stars which approached or will approach the Sun closer than 4.0\,pc. Our list is complete up to a distance of about 2.5\,pc, but includes also some more distant perturbers mentioned in earlier papers, which due to the update of their astrometry increased their minimal distance, including the massive Algol system.
 }
    \label{fig:histogram647}
\end{figure}

With this approach we have obtained a list of 820~objects consisting of 751~single stars and 69~binary or multiple systems which in the past or in the future, could approach the Sun and as a consequence perturb some LPCs' orbits. Our results based on the most up to date astrometry reveal that 714 of our perturbers encountered or will encounter the Sun within a distance smaller than 10\,pc and 647 were or will be closer than 4.0\,pc. We finally accepted this last proximity threshold to allow for the Algol system to be included. It is important because of its large mass of 6\,$M_{\odot}$ and extremely small relative velocity of 2\,km s$^{-1}$. The distribution of the minimal distances between these stars and the sun is presented in Fig.~\ref{fig:histogram647}, and the ten closest past stellar passages are listed in Table~\ref{tab:stars_mindist}.

\begin{table*}

\caption{Ten closest past stellar passages near the Sun obtained in the present research. Star designations, the minimal distance from the Sun, time of the closest approach to the Sun, the relative velocity, and the estimated mass of the star are shown. Two stars for which we detected strong action on the observed long period comets are given in bold.}\label{tab:stars_mindist}
\centering
\begin{threeparttable}[t]
 \begin{tabular}{c c c c c c} 
 \hline
 SIMBAD name & Gaia DR2 name & mindist [pc] & time [Myr] & vrel [km/s] & mass \\ \hline
                         &  Gaia DR2 955098506408767360   &  0.089 &  $-$0.736 &  38.51 &  1.26 \\ 
                         &  Gaia DR2 5571232118090082816  &  0.181 &  $-$1.166 &  82.27 &  0.78 \\ 
                         &  \bf{Gaia DR2 5700273723303646464}\tnote{1}  &  \bf{0.183} &  \bf{$-$1.639} &  \bf{38.08} &  \bf{0.95} \\ 
                         &  Gaia DR2 2946037094755244800  &  0.242 &  $-$0.906 &  42.11 &  0.36 \\
                         &  Gaia DR2 52952724810126208    &  0.255 &   $-$0.540 &  37.77 &  1.46 \\
WISE J072003.20-084651.2 &                                &  0.286 &   $-$0.071 &  83.19 &  0.07 \\
TYC 6552-1735-2          &  Gaia DR2 5599691155509093120  &  0.333 &   $-$0.840 &  75.02 &  1.10 \\
TYC 6487-696-1           &  Gaia DR2 2906805008048914944  &  0.393 &   $-$1.715 &  47.66 &  1.60 \\
\bf{HD 7977}     &  \bf{Gaia DR2 510911618569239040}\tnote{2}   &  \bf{0.406} &  \bf{$-$2.799} &  \bf{26.48} &  \bf{1.10} \\
TYC 5972-2542-1          &  Gaia DR2 2929487348818749824  &  0.454 &  $-$5.423 &  70.00 &  1.34 \\
 \hline
 \end{tabular}
 \begin{tablenotes}
     \item[1] Encounter with comet C/2012~F3 PANSTARRS is discussed in Sec. \ref{sec:2012f3}. 
     \item[2] Encounter with comet C/2002~A3 LINEAR is discussed in Sec. \ref{sec:2002a3}.
   \end{tablenotes}
 \end{threeparttable}
\end{table*}

\section{Increasing set of precise Oort spike orbits}\label{Increasing_LPCs_sample}

For decades the Catalogue of Cometary Orbits prepared by Brian Marsden and his collaborators remained an indispensable source of LPCs' orbital data. After its last edition \citep{marsden-cat:2008} the sample of precise orbits must be gathered from different sources. In a series of papers (accompanied with several publicly available catalogues), some of us are involved in, calculations or recalculations of a great number of LPC orbits are presented  
\citep[][and references therein]{krol-sit-et-al:2014,krolikowska:2014,kroli-dyb:2016,kroli_dyb:2017,Kroli-Dyb:2018}.

\subsection{Investigated sample of Oort spike comets and their orbits}\label{sec:LPCs_sample}

Using these sources we investigate here a sample of 277~long period comets having an original $1/a_{\rm ori} <0.000100$\,au$^{-1}$ (in some individual cases this limit is shifted to the value of $0.000200$\,au$^{-1}$). This is almost a complete sample of Oort spike comets discovered in years 1901--2012, with an inclusion of large-perihelion comets ($q>3.1$\,au) extended to 2017, see \cite{kroli-dyb:2019}. The orbits of numerous comets were already published in the papers and catalogues quoted above, however many of them are still unpublished, and will be included in our two forthcoming  publications (Dybczyński and Królikowska, in preparation).

For many of them non-gravitational (hereafter NG) orbits were determinable but we included in our sample of cometary orbits also their purely gravitational (hereafter GR) orbits to test widely a possibility of close encounters with known stars. Moreover, for many small-perihelion comets with long data-arcs it was possible to obtain NG/GR~orbits using some subset of data (for example pre-perihelion arc of data or arc of observations taken at large distances).
Therefore, sometimes several orbital solutions are offered for a single comet, including one that is the preferred orbit for past dynamical long-term calculations.

In this way we tested in total 498~different orbital solutions for 277~comets to search for stellar perturbations in their past dynamical evolution. 

\begin{figure*}
    \centering
    \includegraphics[width=0.47\linewidth]{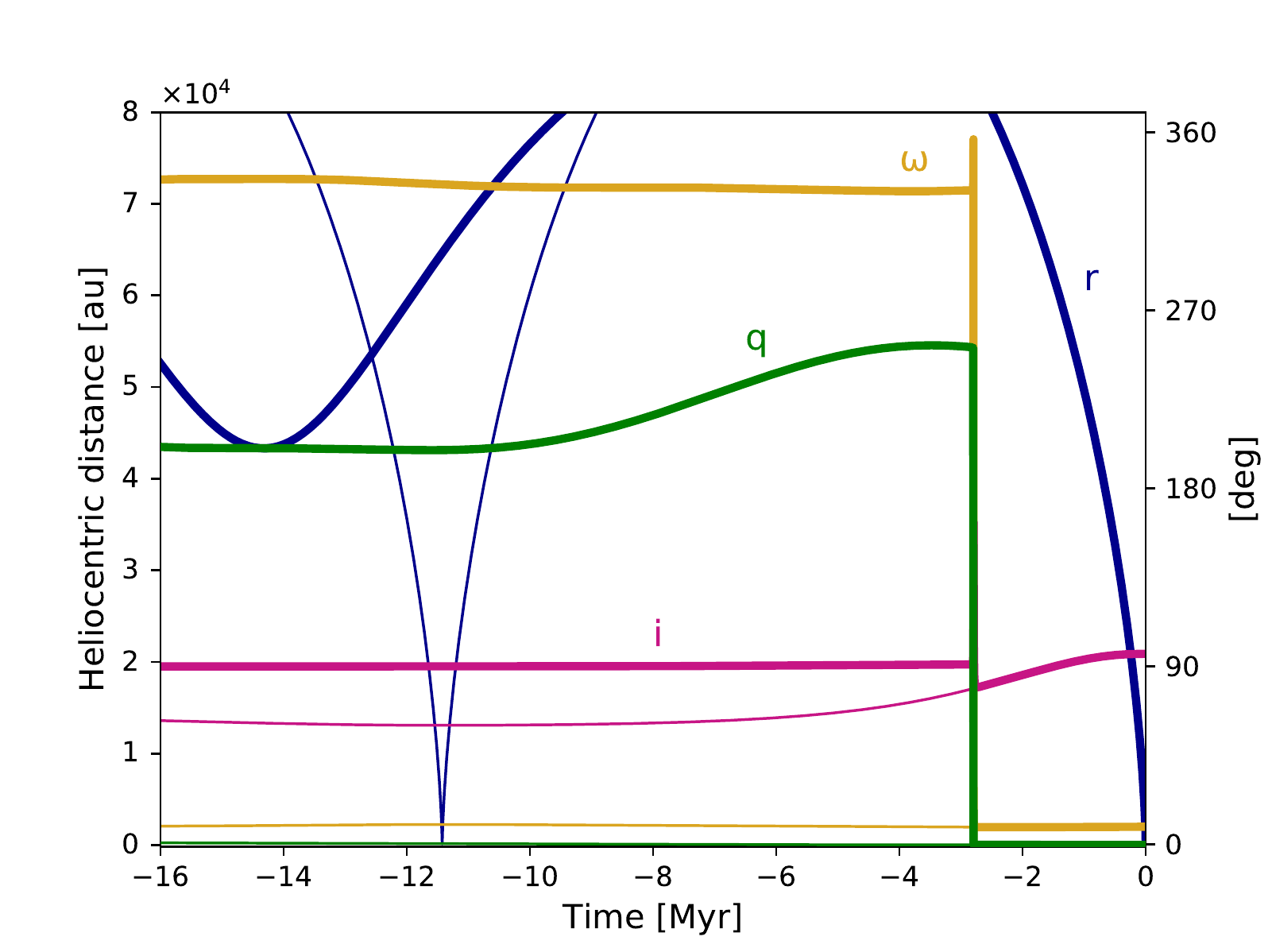}
    \includegraphics[width=0.47\linewidth]{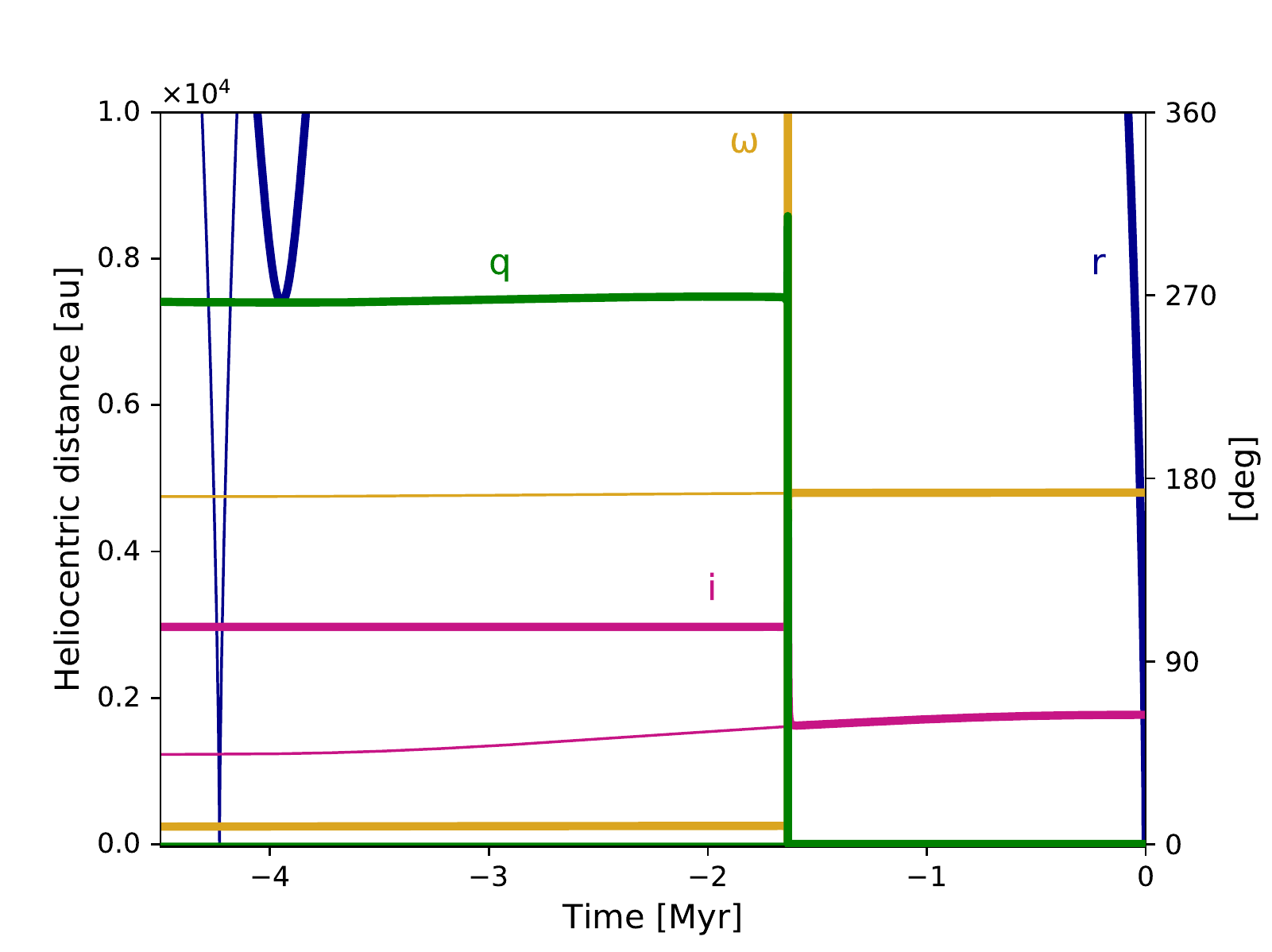}
    \caption{Past evolution of the nominal orbit of C/2002~A3 (left panel) and C/2012~F3 (right panel)  under the simultaneous Galactic and stellar perturbations. See text for a detailed explanation. }
    \label{fig:2002a3_0230_2012f3_0507}
\end{figure*}

\section{Two examples of a possible strong stellar action on comets}\label{sect:two_examples}

After completing the final list of possible stellar perturbers we were able to search for close approaches of all candidate stars with all comets at hand. The very first step was to integrate numerically each cometary orbit with each stellar perturber backwards as a 3-body system (the Sun -- comet -- star) under the influence of the Galactic potential. 

It appeared, that for the first time, we detected  strong star -- comet interactions. In two cases the miss distance is at the level of a few hundred au and the resulting changes in cometary orbit are spectacular. It should be noted that these two cases are found by a quick search using only nominal cometary orbits and nominal stellar data, more detailed inspection is in progress. 
For these two cases more thorough calculations, involving thousands of clones of both the comet and the star, were performed.  Below both these cases are described in detail.

\section{The case of C/2002 A3 LINEAR}\label{sec:2002a3}

Comet C/2002~A3 was discovered on January 13, 2002 by LINEAR as an apparently asteroidal object at a distance of 5.21\,au from the Sun, one day after its closest approach to Earth (4.395\,au).  Soon, its cometary appearance was reported \citep[see][]{IAUC-7799}. The comet reached its perihelion (5.15\,au from the Sun) on April 24, 2002, and next was followed for more than one year up to June 24, 2003 (6.05\,au from the Sun). Around its perihelion the comet was not observed for almost 7.5 months (from April 2, 2002 to November 10, 2002).  

The osculating heliocentric orbit (for the epoch close to the perihelion passage)  was obtained using 294 positional observations spanning over 1.44\,yr. The orbit is of 1a~class; NG~effects are not detectable in the motion of this comet. 

This osculating orbit was cloned using its covariance matrix (see \cite{kroli-dyb:2010} for more details on the methods used) and 5001\,VCs (virtual comets) were next followed to the past to obtain a set of original barycentric orbits (250\,au from the Sun), including a nominal orbit. This procedure allows us to determine the uncertainties of the orbital elements at each stage of our calculations. The original barycentric orbit is given in Table~\ref{tab:2012f3_2002a3_models}.

This is a unique comet from a dynamical point of view because it experienced strong planetary perturbations during its journey through the planetary zone; it had a close encounter with Jupiter on January 22, 2003 (miss distance of 0.502\,au). Planetary perturbations changed the orbit of C/2002~A3 significantly, and the comet will leave the planetary zone in an orbit with semimajor axis of only about 160\,au. It means that its orbital period has been shortened to about 2000\,yr. 

Our orbital solution (GR b5, see Table \ref{tab:2012f3_2002a3_models}) showed a violent past dynamics as a result of a close encounter with the star  HD\,7977.

\begin{table}
    \centering
    \caption{Stellar astrometric and kinematic data obtained from the Gaia DR2 catalogue and the adopted masses. These values, together with the corresponding covariance matrices, were used for the calculations presented in Sec. \ref{sec:2002a3} and \ref{sec:2012f3}.}
    \begin{tabular}{l r r} \hline
         & & \textbf{Gaia DR2}\\
         &  \textbf{HD 7977}  &  \textbf{5700273723303646464} \\
         \hline
    
       $\alpha [$deg$]$  & 20.1316471 & 124.3038106 \\
       
       $\sigma_{\alpha^*} [$mas$]$  & 0.0229  & 0.7028\\
       &&\\
       $\delta [$deg$] $  & 61.88252161 & $-$21.78914448 \\
     
       $\sigma_{\delta} [$mas$] $  & 0.0261 & 0.7006\\
       &&\\
       $\pi [$mas$]$ & 13.2030 &  15.6697 \\
              & $\pm$0.0376 & $\pm$1.0835\\
              &&\\
       $\mu_{\alpha^{*}} [$mas/yr$] $  & 0.559  & 0.155 \\
                            & $\pm$0.040 & $\pm$ 1.389 \\
                            &&\\
       $\mu_{\delta} [$mas/yr$]$  & 0.014 & $-$0.213\\
                     &  $\pm$0.046     & $\pm$1.355\\
                     &&\\
       $v_{r} [$km/s$]$  & 26.45  & 38.05\\
       &$\pm$0.35 & $\pm$0.91\\
       &&\\
       mass [$M_\odot$]  & 1.10 & 0.95 \\
       \hline
    \end{tabular}
    
    \label{tab:stars}
\end{table}

\subsection{The influence of the star--comet encounter on the nominal orbit of C/2002~A3}\label{sec:2002A3_star_comet}

We found that the star HD\,7977, designated also as BD+61~250 or TYC~4034-1077-1, had a big impact on C/2002~A3's past motion. The star can be found in the Gaia~DR2 catalogue as the object DR2~510911618569239040. All five astrometric parameters can be found in DR2, as well as the radial velocity (see Table~\ref{tab:stars}). Various quality parameters show that the astrometric accuracy is good, this star was designated as a 'primary astrometric star'.

As concerns its astrophysical parameters, HD\,7977 is most often described in the literature as a main sequence dwarf of the spectral type G0, see for example \cite{Buscombe:1998}. In Gaia~DR2 there are all parameters relevant for mass estimation, i.e. an effective temperature, a stellar radius  and a luminosity. Our various mass estimates as well as some found in the literature are rather consistent and we adopted 1.10\,M$_{\odot}$ as its mass.

In the absence of any stellar perturbations comet C/2002~A3 can be traced numerically under the influence of the Galactic potential back to its previous perihelion passage  about 11.5\,Myr ago. Its perihelion distance changes from the observed 5.14\,au up to over 126\,au and its eccentricity changes from 0.9999 to 0.9975. After including the star HD\,7977 into the dynamical model we were surprised to obtain a nominal previous perihelion distance of over 43\,000 au and a previous eccentricity as small as 0.25! The orbital period increased to over 14\,Myr. All the above results are for the nominal comet orbit and the nominal stellar data. This dynamical evolution is shown in the left panel in Fig.~\ref{fig:2002a3_0230_2012f3_0507}. 

\begin{figure*}
    \centering
    \includegraphics[width=0.47\linewidth]{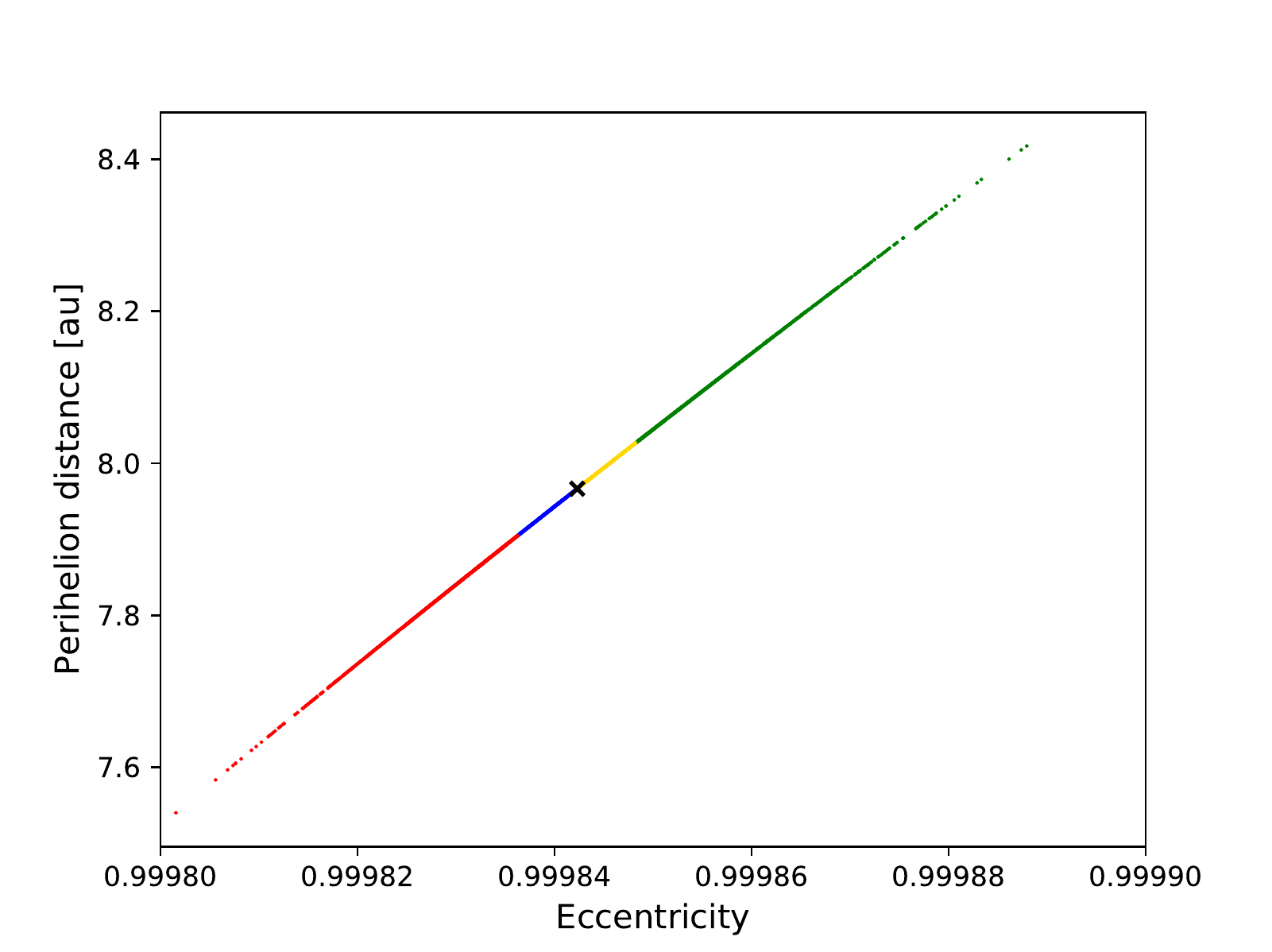}
    \includegraphics[width=0.47\linewidth]{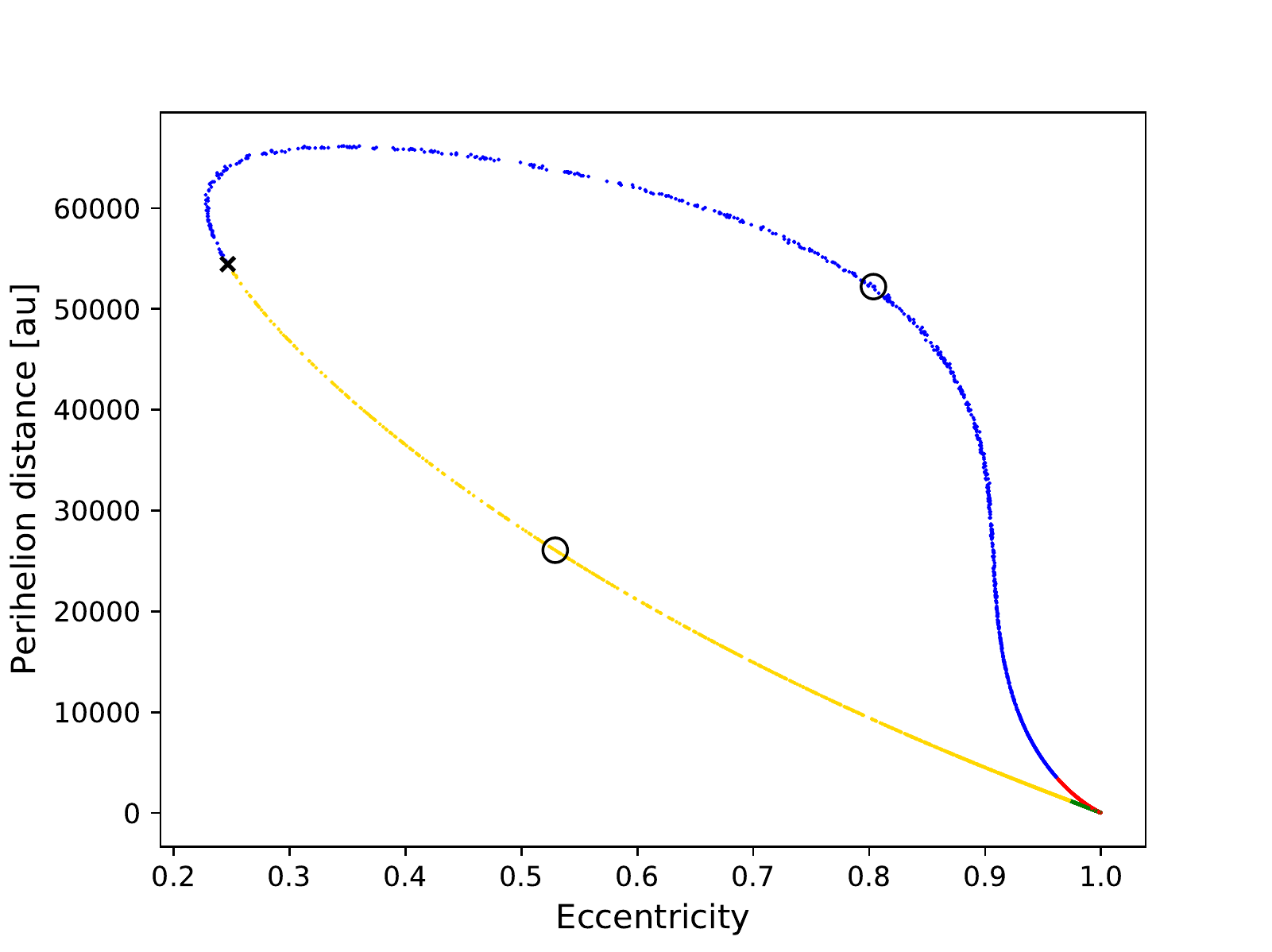}
    \caption{The $q$ -- $e$ distribution of 5001~VCs of C/2002~A3 
    2\,Myr ago, left panel and 3\,Myr ago, right panel. A black cross marks the nominal comet orbit and its full past evolution is presented in the left panel of Fig.~\ref{fig:2002a3_0230_2012f3_0507}.  VCs are consistently coloured in both plots to show how they substantially evolved due to the close encounter with this particular star. Left panel: the eccentricity at this moment still has a Gaussian distribution with $\sigma = 0.0000093$, and the yellow and blue clones differ from the nominal value by less than $0.5 \sigma$. Right panel:  One can easily note that the $1\sigma$ interval of yellow and blue clones from the left panel now occupies a really huge area in this plot. In the right panel clones further examined in Fig. \ref{fig:clonesevolution} are marked with circles, in the left panel they are too close to the nominal orbit to be distinguished in
this manner.}
    \label{fig:qe_2002a3_2Myr_3Myr}
\end{figure*}

In this picture, the horizontal time axis extends from the observed perihelion passage (time=0) back to the previous one. The left vertical axis is expressed in au and corresponds to the osculating perihelion distance plot ($q$, green lines) as
well as the heliocentric distance plot ($r$, blue lines).  The right vertical axis is expressed in degrees and describes the evolution of the osculating inclination ($i$, magenta lines) and the argument of perihelion ($\omega$, yellow lines). Both these angular elements are expressed in the Galactic frame. For each orbital parameter its evolution in the absence of the stellar action is depicted with thin lines while the stellar-driven dynamics is described with thick lines.

Such a star--comet interaction ideally corresponds to the scenario proposed by \cite{oort:1950}: a passing star acts on a comet residing in the Oort cloud having moderate eccentricity (here 0.25) and changes its orbit to an observed, near-parabolic one. This spectacular orbit change is possible because of a very close stellar passage: we calculated that the minimum star -- comet distance for the nominal orbit was 740\,au. It should be reminded here that these numbers were obtained for the nominal comet orbit and the nominal stellar data. In both cases we have to deal with their uncertainties, which we describe in the following sections.

\subsection{The influence of the star on the swarm of cometary orbits}\label{sec:2002A3_star_VCs}

As we stated in Section~\ref{sec:2002a3}, just after obtaining an osculating comet orbit we generate 5\,000 of its clones, called virtual comets. As a result we have also 5\,001 different original orbits (a nominal one plus 5\,000~VCs) reflecting the uncertainty in orbital elements at the stage of orbit determination from the positional data. 
To observe how the uncertainty in the comet orbit influences the results of the reconstructed strong stellar perturbation we numerically integrate all 5\,001~VCs as 3-body cases under the Galactic potential. In Fig.~\ref{fig:qe_2002a3_2Myr_3Myr} we present two snapshots of the VCs' swarm distribution  using the $q$ -- $e$ plane taken 2\,Myr and 3\,Myr ago. The closest star--comet approach took place 2.8\,Myr ago.

\begin{figure*}
    \centering
    \includegraphics[width=0.47\linewidth]{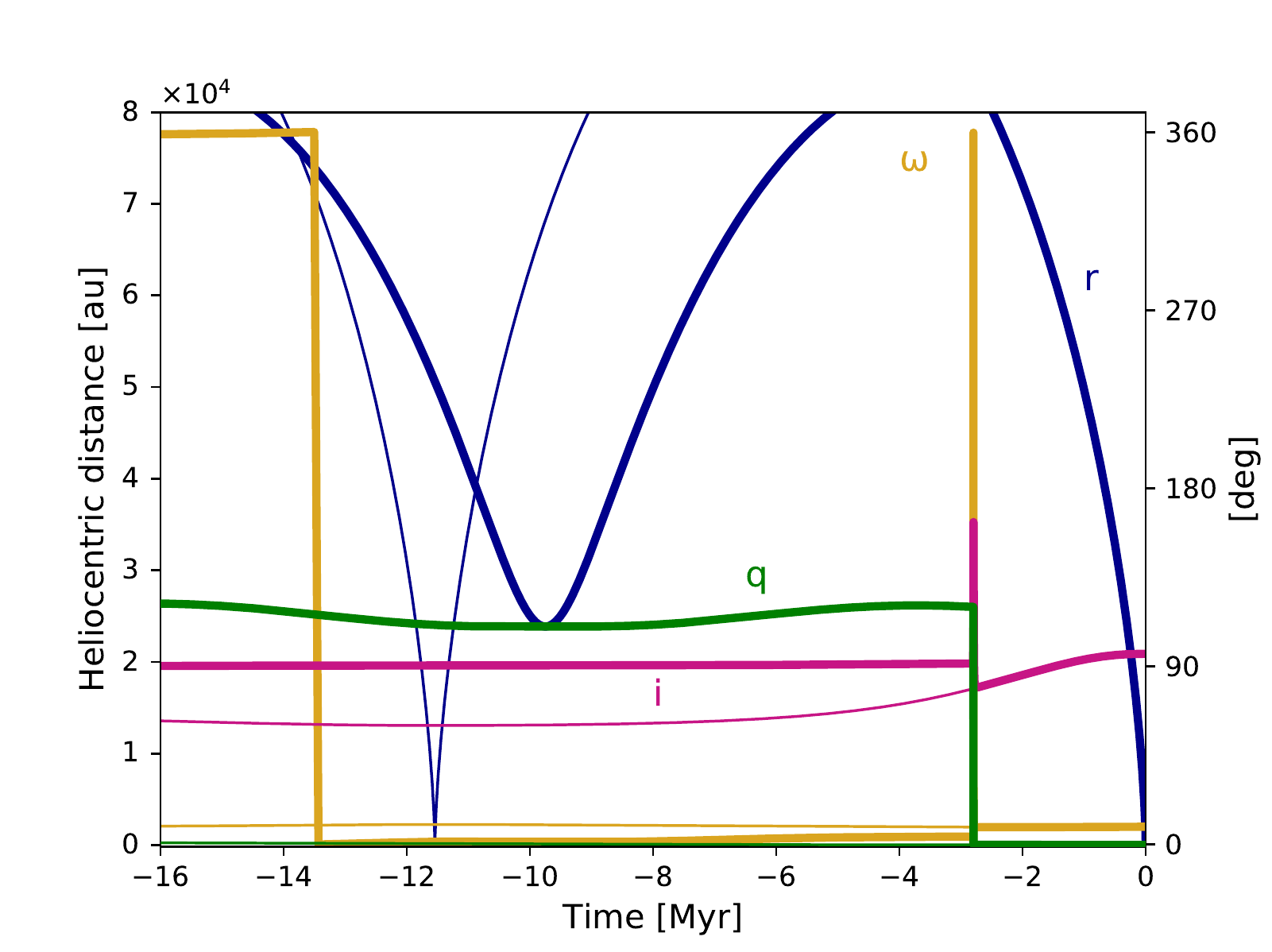}
    \includegraphics[width=0.47\linewidth]{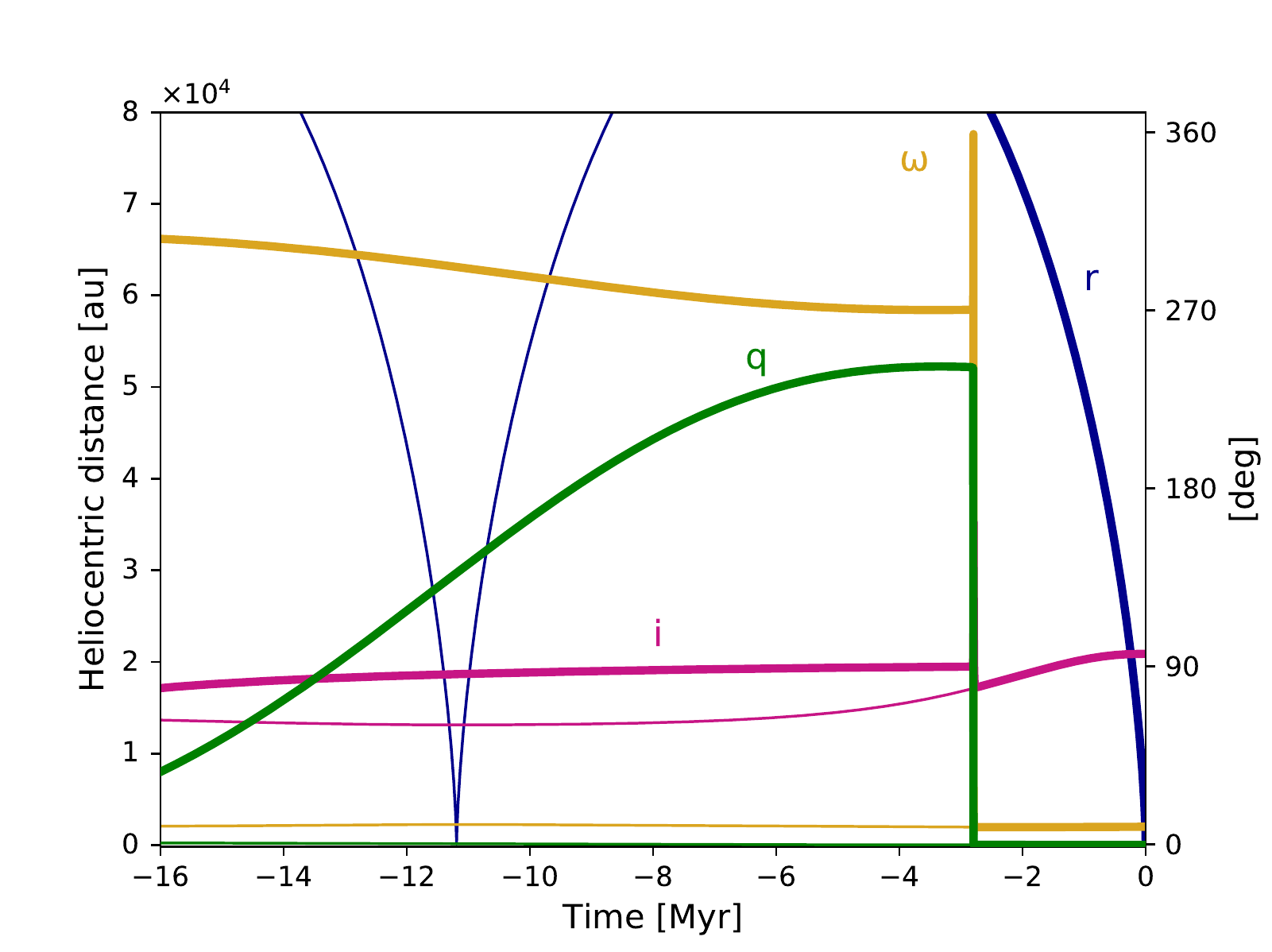}
    \caption{Past evolution of the selected yellow (left panel) and blue (right panel) clone (marked with circles in the right panel of Fig. \ref{fig:qe_2002a3_2Myr_3Myr}) of C/2002~A3 under the simultaneous Galactic and stellar perturbations. See text in Sec. \ref{sec:2002A3_star_comet} for a detailed explanation. }
    \label{fig:clonesevolution}
\end{figure*}

\begin{figure}
    \centering
    \includegraphics[width=\columnwidth]{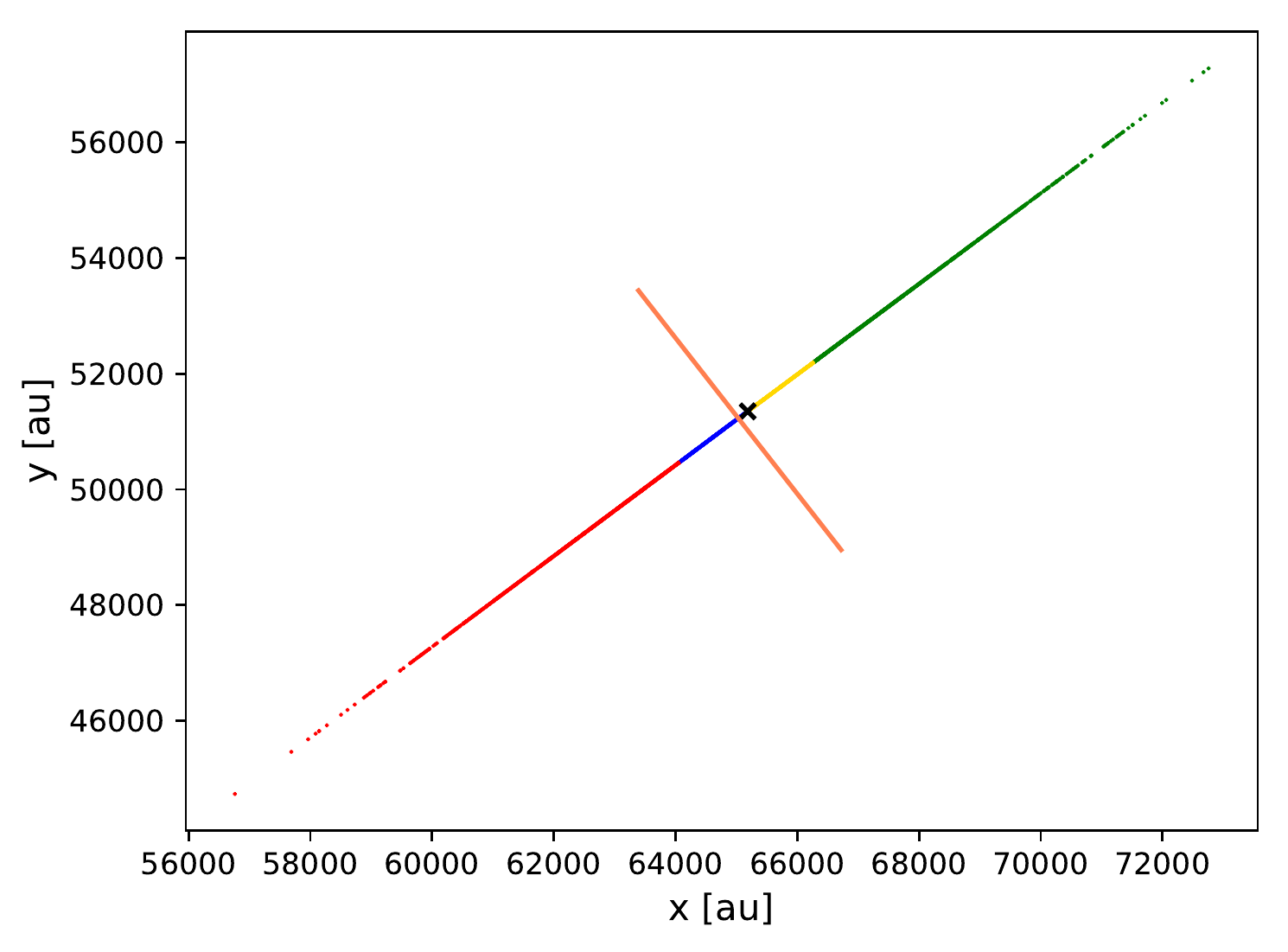}
    \caption{The geometry of  VCs swarm of C/2002~A3 with respect to the stellar path, projected onto the Galactic disk plane. The star trajectory in 3D is about 670\,au above the line of comet clones. All comet clones are stopped at the moment of the closest proximity of the comet nominal to the star. VCs are coloured in the same manner as in Fig.~\ref{fig:qe_2002a3_2Myr_3Myr}. The orange line depicts one thousand years of the star movement. }
    \label{fig:2002a3_0230_geometry}
\end{figure}

\begin{figure*}
    \centering
    \includegraphics[width=0.47\linewidth]{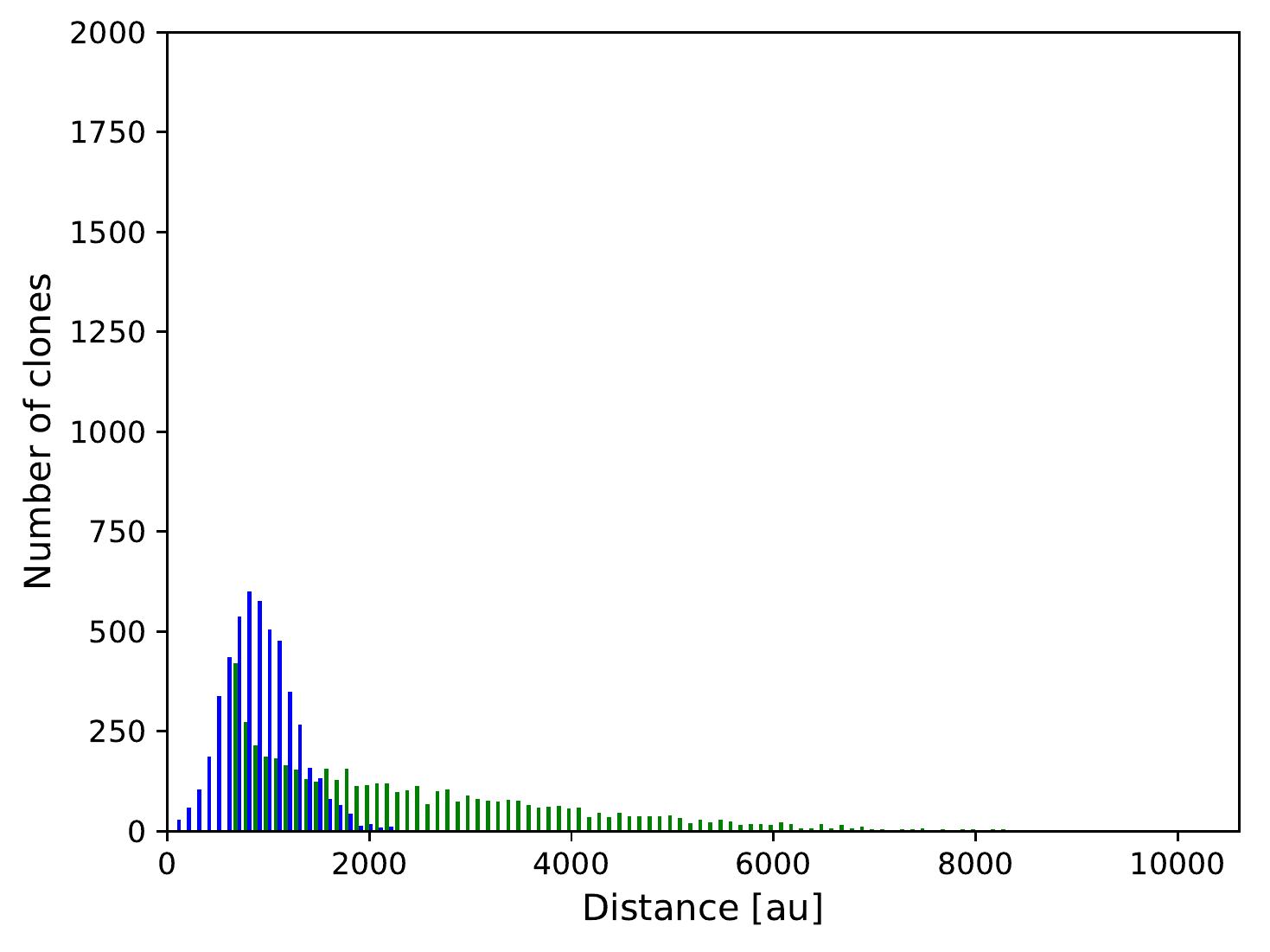}
    \includegraphics[width=0.47\linewidth]{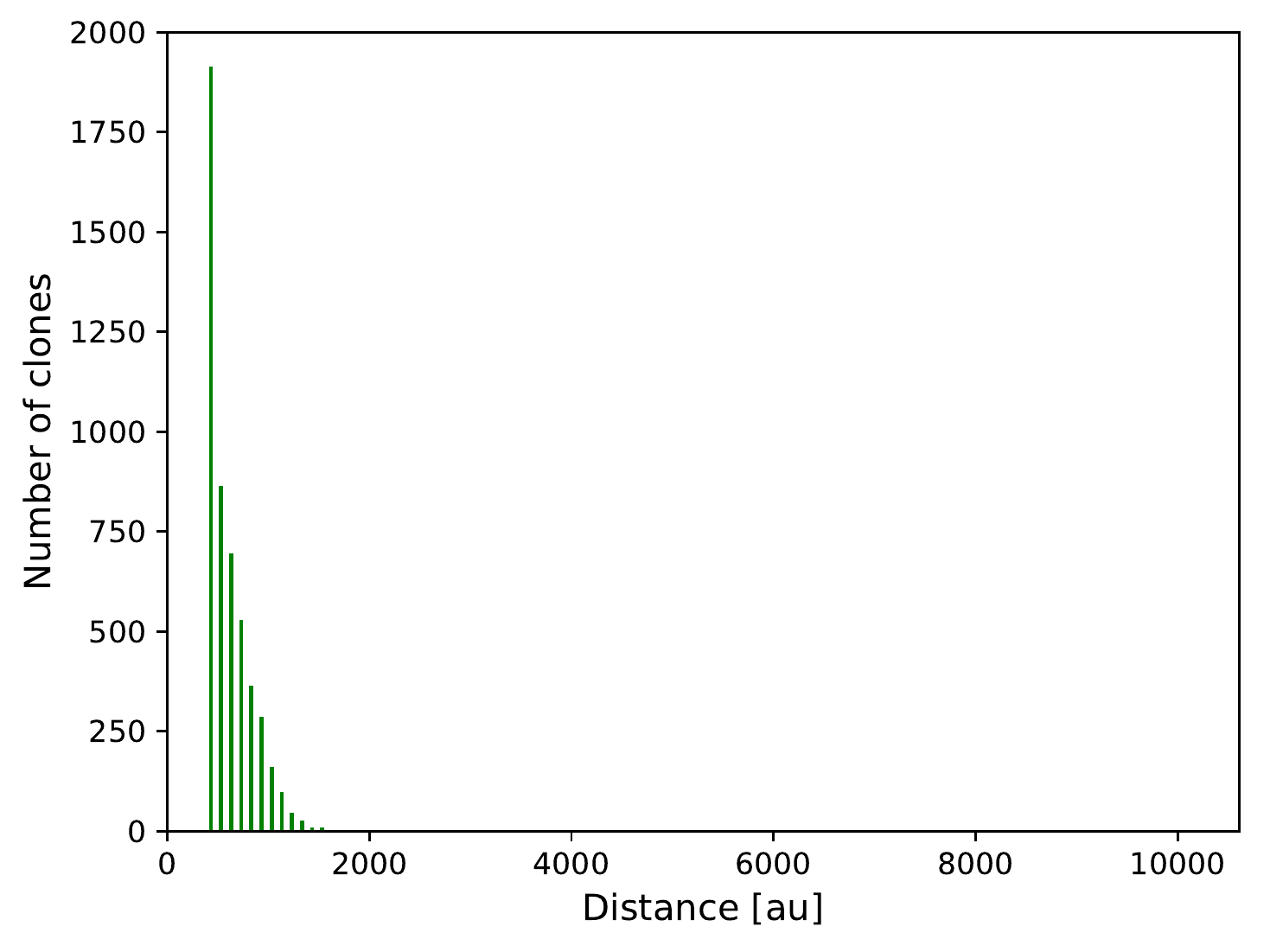}
    \caption{Left panel: Distributions of minimal distances between the clones of C/2002~A3  and the HD~7977 star clones. Blue bars indicate distances between the nominal cometary orbit and the clones of the star, whereas green bars designate distances of comet clones from the nominal star path. The minimal distance for the nominal pair equals 740\,au. Right panel: Distribution of minimal distances between nominal Star~B and clones of C/2012\,F3. For the nominal comet the distance equals 553\,au.}
    \label{fig:histograms}
\end{figure*}

In the left panel in Fig.~\ref{fig:qe_2002a3_2Myr_3Myr} we see a typical $q$--$e$ distribution of VCs obtained from the original comet orbit. This distribution is compact for such a high quality comet orbit and evolves slowly during the past motion under the Galactic tides. The situation drastically changes as a result of the strong interaction with the star, as shown in the right panel of Fig.~\ref{fig:qe_2002a3_2Myr_3Myr}. The ranges of the perihelion distance and the eccentricity are really large due to this close encounter with the star. The interpretation of the 'loop shape' of this plot is as follows. As it is shown in Fig.~\ref{fig:2002a3_0230_geometry} a star moves almost perpendicular to the swarm of VCs and its closest proximity (less than 700\,au) is near the nominal orbit (depicted with a black cross). VCs near the nominal are therefore perturbed very strongly so their eccentricities vary from smaller to bigger depending on the relative geometry and the distance to the star. Clones from the wings of the swarm (green and red) are perturbed much less and most of them are simply concentrated in a small region at the bottom right corner of the right panel in Fig.~\ref{fig:qe_2002a3_2Myr_3Myr} due to the huge range of $q$ in this panel. One can observe that yellow and blue clones, which are very close to the nominal in the left panel of Fig.~\ref{fig:qe_2002a3_2Myr_3Myr} (in phase space) and in Fig.~\ref{fig:2002a3_0230_geometry} (in real space) are in the right panel of Fig.~\ref{fig:qe_2002a3_2Myr_3Myr} smeared out at large ranges of $q$ and $e$. But their minimal distances from the star differ not so much and the relative geometries are similar. This wide distribution reflects the large (and obvious!) sensitivity of the result of the star--comet  approach on the minimal distance and the relative geometry of the star passage. In Fig.~\ref{fig:clonesevolution} we plotted the evolution of orbital elements of two clones selected from yellow and blue ones, marked with circles in Fig. \ref{fig:qe_2002a3_2Myr_3Myr}. This  presents how differently clones evolve under the influence of HD\,7977. These evolution plots fully explain the different positions of these clones in Fig. \ref{fig:qe_2002a3_2Myr_3Myr}. Red and green clones generally do not experience strong stellar perturbations and
their dynamical evolution is very similar to that depicted with thin lines.

It has to be stressed here that the dynamical evolution of the VCs' swarm {\it  is described above in the reversed time}. In reality comet C/2002~A3 approached HD\,7977 in one particular orbit from among thousands depicted as dots in the right panel of Fig.~\ref{fig:qe_2002a3_2Myr_3Myr} and than evolved due to the star -- comet interaction into the corresponding orbit, i.e. a single dot, in the left panel of this figure, and then (in 2002) the comet was observed.

\subsection{Star--comet encounter and stellar data uncertainties}\label{sec:2002A3_star_comet_full}

We can study the influence of stellar data uncertainties in a similar way. To this aim we generated 5\,000~clones of the discussed star. Clones were drawn using a multivariate normal distribution with the help of covariance matrix elements which are included in the Gaia DR2 catalogue. We chose to follow this approach because it was proven that the use of a covariance matrix improves its relevance (for discussion on this topic and the method of drawing clones see for example \cite{berski-dyb:2016}). The radial velocity was drawn independently from the assumed Gaussian distribution. Because of its unknown uncertainty we decided not to draw the stellar mass. To examine the star parameters uncertainty influence we numerically integrated the comet's nominal orbit with the nominal star and its 5\,000 clones, each time as a 3-body problem under the Galactic potential. As for cometary clones the integration was stopped when the comet reached its previous perihelion or its heliocentric distance exceeded 0.6\,pc. In the left panel of  Fig.~\ref{fig:histograms} we compare the influence of the cometary orbit uncertainty and that of the stellar data by showing the distributions of the minimal star -- comet distances. It can be observed that the blue histogram, which presents minimal distances of star clones from the comet's nominal orbit, is much more compact than the green histogram, showing the distribution of the minimal distances between the nominal star path and comet clones.

In some sense we can conclude that in this particular case, our knowledge of the distant motion of the star is better than that of the comet, which is the result of the high precision data in the Gaia DR2 catalogue.

As concerns the dynamical history of comet C/2002~A3, based on Fig.~\ref{fig:qe_2002a3_2Myr_3Myr} we can state that, before they met star HD~7977, most of the cometary clones had their perihelion distance smaller than  5\,000\,au. It seems that a lot of clones are in the upper part of the right panel of Fig. \ref{fig:qe_2002a3_2Myr_3Myr}. However, our VCs' swarm consists of 1513 green clones and 1554 red clones and they (more than 60 per cent of all clones) are densely packed in the right bottom corner of the plot. On the other hand, in the upper part of it we have 1414 clones, including the nominal orbit, having their perihelion distance greater than 5\,000\,au. The comet orbit could be elliptical or near-parabolic, as most of the clones have their eccentricity close to 1.0. Whichever variant is true, this example seems to be an  unique case of the orbital evolution during one passage through the inner Solar System, because this comet was also caught in a very tight orbit (future semimajor axis of about 160\,au), under the influence of planetary perturbations (see also beginning of this section). 

In short, our planetary system caught an Oort cloud comet, or a hyperbolic one, with the evident help of HD\,7977. In both cases it was a new comet for sure.

\section{The case of C/2012~F3 PANSTARRS}\label{sec:2012f3}

Comet C/2012~F3 was discovered by the Pan-STARRS\,1 telescope (Haleakala) on March 16, 2012 at a distance of 9.57\,au from the Sun. 
The comet reached its perihelion (3.46\,au from the Sun) on April 6, 2015, and was followed to October 17, 2017 (8.36\,au from the Sun).
A few prediscovery images were obtained on January 19, 2012 (9.93\,au from the Sun) and March 2, 2012, extending the data arc to 5.74\,yr (1763~positional measurements in total). Such a long data arc allowed us to study its orbit in greater detail by, among others, obtaining a NG orbit of a superior quality (1a+ class) \cite[for quality class and method of orbit determination see][]{kroli-dyb:2013,kroli_dyb:2017}. When it turned out that this comet might have experienced a close stellar approach in its past, we updated our calculations using the positional data set retrieved from the Minor Planet Center in June~2019, but new solutions lead to the same conclusions.

Since the perihelion distance is moderately large we checked here, as in \cite{kroli_dyb:2017}, two different formulas for a $g(r)$-like function, that is, the standard $g(r)$ describing water ice sublimation (solution l5) and $g(r)$-like with the distance scale parameter $r_0$=10\,au, more adequate for CO~sublimation (solution d5). Both these NG~solutions fit the data with RMS  of 0\farcs 34, whereas the purely gravitational model of motion (solution b5) results in RMS of 0\farcs 55. Table~\ref{tab:2012f3_2002a3_models} presents these three solutions for the original orbit of this comet (250\,au from the Sun and before entering the planetary zone). The preferred original orbit is that based on the $g(r)$-like function with $r_0$=10\,au and with weighting procedure included in the orbit determination process (solution d5, given as the last one in the table).

\begin{table*}
	\caption{Original barycentric orbits of C/2002~A3 (part [A]) and C/2012~F3 (part [B]) based on full positional data set spanning 1.44\,yr and 5.74\,yr, respectively,  available at the Minor Planet Center on June 24, 2019. Equator and ecliptic of J2000 is used.  Preferred orbit for C/2012~F3 is the NG,~d5 solution. }
	\label{tab:2012f3_2002a3_models}
	{\small{
			\setlength{\tabcolsep}{6.5pt} 
			\begin{tabular}{ccrrrrrrr}
				\hline 
&&&&&&&& \\
\multicolumn{9}{c}{\bf [A]~~~C/2002~A3 LINEAR; ~~~~purely gravitational orbit} \\
&&&&&&&& \\
model  &  Epoch    &  T                 &    $q$[au]    &    $e$        &  $\omega$   &  $\Omega$   &   $i$       & $1/a_{\rm ori}$ \\
&&&&&&&& \\
GR, b5 &  16960906 &  20020425.19669560 &    5.14305567 &    0.99989809 &  329.552533 &  136.668877 &   48.065977 &    19.82 \\
&           &     $\pm$0.00314686 &$\pm$.00000753 &$\pm$.00000925 &$\pm$.000351 &$\pm$.000010 &$\pm$.000016 &$\pm$     1.80 \\
&&&&&&&& \\
\multicolumn{9}{c}{\bf [B]~~~C/2012~F3 PANSTARRS} \\
&&&&&&&& \\
model  &  Epoch    &  T                 &    $q$[au]    &    $e$        &  $\omega$   &  $\Omega$   &   $i$       & $1/a_{\rm ori}$ \\
&&&&&&&& \\
\multicolumn{9}{c}{\bf Purely gravitational original orbit} \\
GR, b5 &  17120506 &  20150407.37226040 &    3.45104212 &    0.99987545 &  104.138424 &  164.664046 &   11.354249 &    36.09 \\
&           &     $\pm$0.00013808 &$\pm$.00000066 &$\pm$.00000166 &$\pm$.000033 &$\pm$.000021 &$\pm$.000004 &$\pm$     0.48 \\
&&&&&&&& \\
\multicolumn{9}{c}{\bf Original NG~orbit obtained using standard $g(r)$ function} \\
NG, l5 &  17120506 &  20150407.38806504 &    3.45096031 &    0.99985170 &  104.142858 &  164.662828 &   11.354037 &    42.97 \\
&           &     $\pm$0.00023159 &$\pm$.00000134 &$\pm$.00000140 &$\pm$.000051 &$\pm$.000023 &$\pm$.000004 &$\pm$     0.41 \\
&&&&&&&& \\
\multicolumn{9}{c}{\bf Original NG~orbit obtained using $g(r)$-like function adequate for CO~sublimation} \\
NG, d5 &  17120506 &  20150407.37848850 &   3.45088127 &    0.99986787 &  104.144170 &  164.662159 &  11.353683 &    38.29 \\
&           &    $\pm$0.00038241 &$\pm$.00000315 &$\pm$.00000199 &$\pm$.000085 &$\pm$.000030 &$\pm$.000008 &$\pm$     0.58 \\
\hline 
\end{tabular}
}}
\end{table*}

\subsection{The influence of the star--comet encounter on the nominal orbit of C/2012~F3}\label{sec:2012F3_star_comet}

For this comet we also detected a strong stellar perturbation in its motion which happened about 1.5 Myr ago.

The star in question is designated as Gaia DR2 5700273723303646464 (hereafter Star~B for short) and its parallax was determined for the first time by the Gaia mission. This stellar data are of much lower quality than those for the previous star (but all five astrometric parameters as well as the radial velocity are published in Gaia DR2, see Table~\ref{tab:stars}). The effective temperature is also given but with the large error, while the stellar radius and the luminosity are missing. This star has not been cross-matched to any other catalogue, either by the Gaia team nor the CDS team\footnote{Centre de Données astronomiques, Strasbourg, France}. Moreover its spectral type and luminosity class cannot be found in the literature. Therefore, for the present calculations we adopted its mass to be equal to 0.95\,M$_{\odot}$ as estimated in \cite{Bailer-Jones:2018}.

\begin{figure*}
    \centering
    \includegraphics[width=0.47\linewidth]{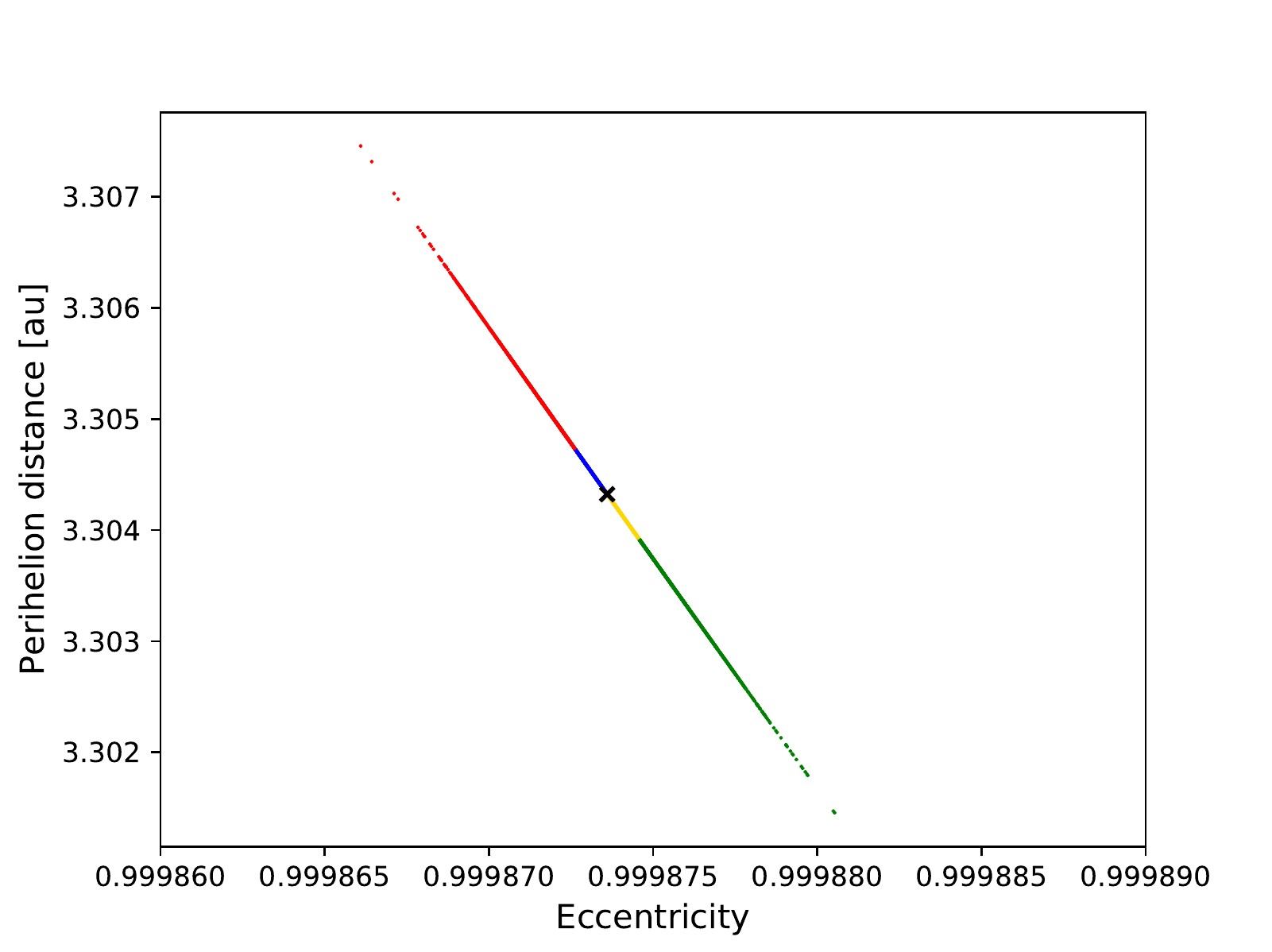}
    \includegraphics[width=0.47\linewidth]{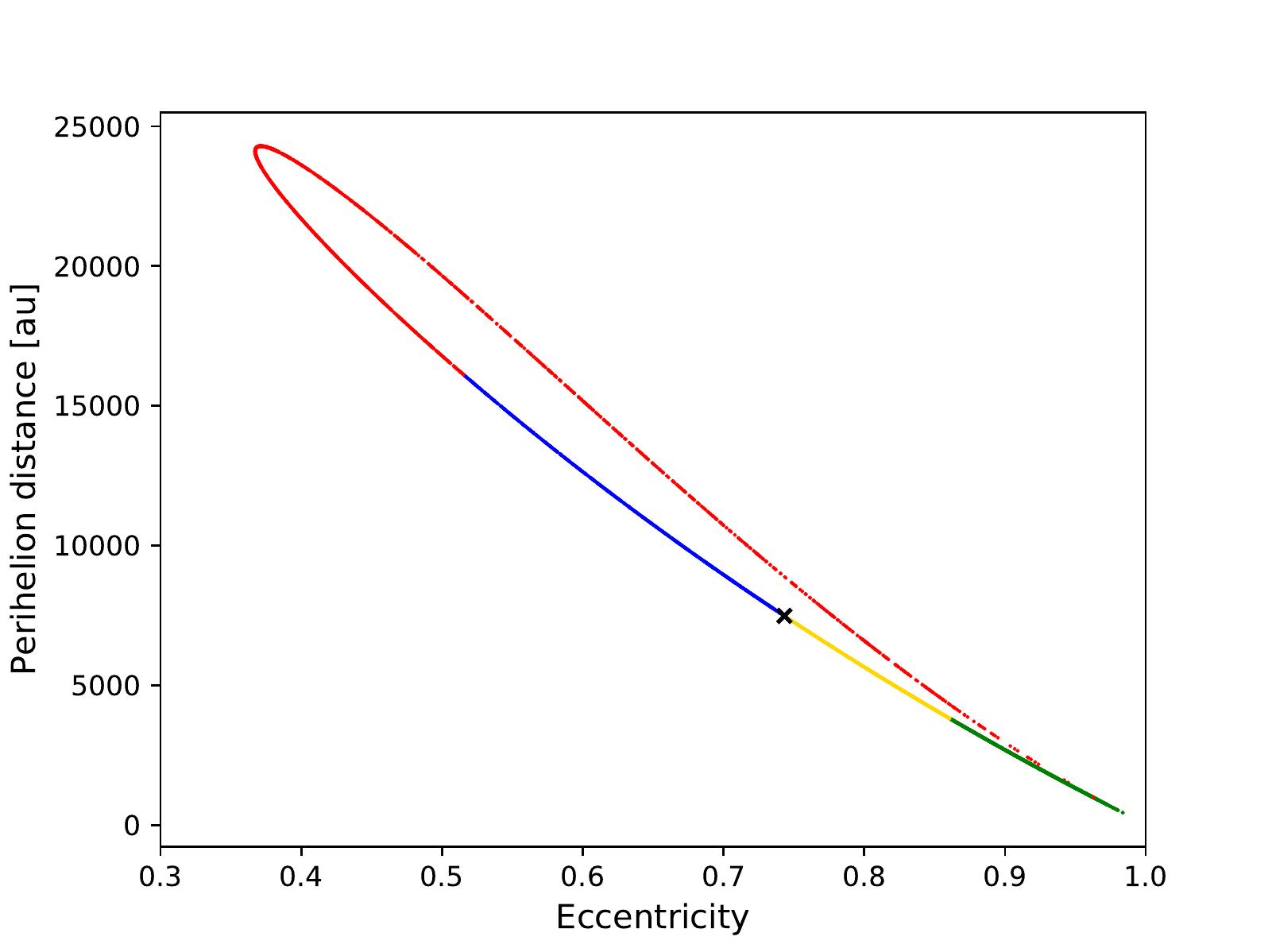}
    \caption{The $q$ -- $e$ distribution of 5\,001~VCs of C/2012~F3 captured 1\,Myr ago (left panel) and 2\,Myr ago (right panel). The closest encounter with the star was 1.64\,Myr ago. A black cross marks a nominal comet orbit. VCs are coloured to show how they evolved in this period. Left panel: The eccentricity at this stage of evolution have a Gaussian distribution with $\sigma = 0.0000020$. The yellow and blue clones differ from the nominal value by at most $0.5 \sigma$. Right panel: One can easily note that the $1 \sigma$ interval of yellow and blue clones from the left panel now occupy much bigger area in this $q$--$e$ plot but smaller than that occupied by red clones. }
    \label{fig:qe_2012f3_1Myr_2Myr}
\end{figure*}

\begin{figure}
    \centering
    \includegraphics[width=\columnwidth]{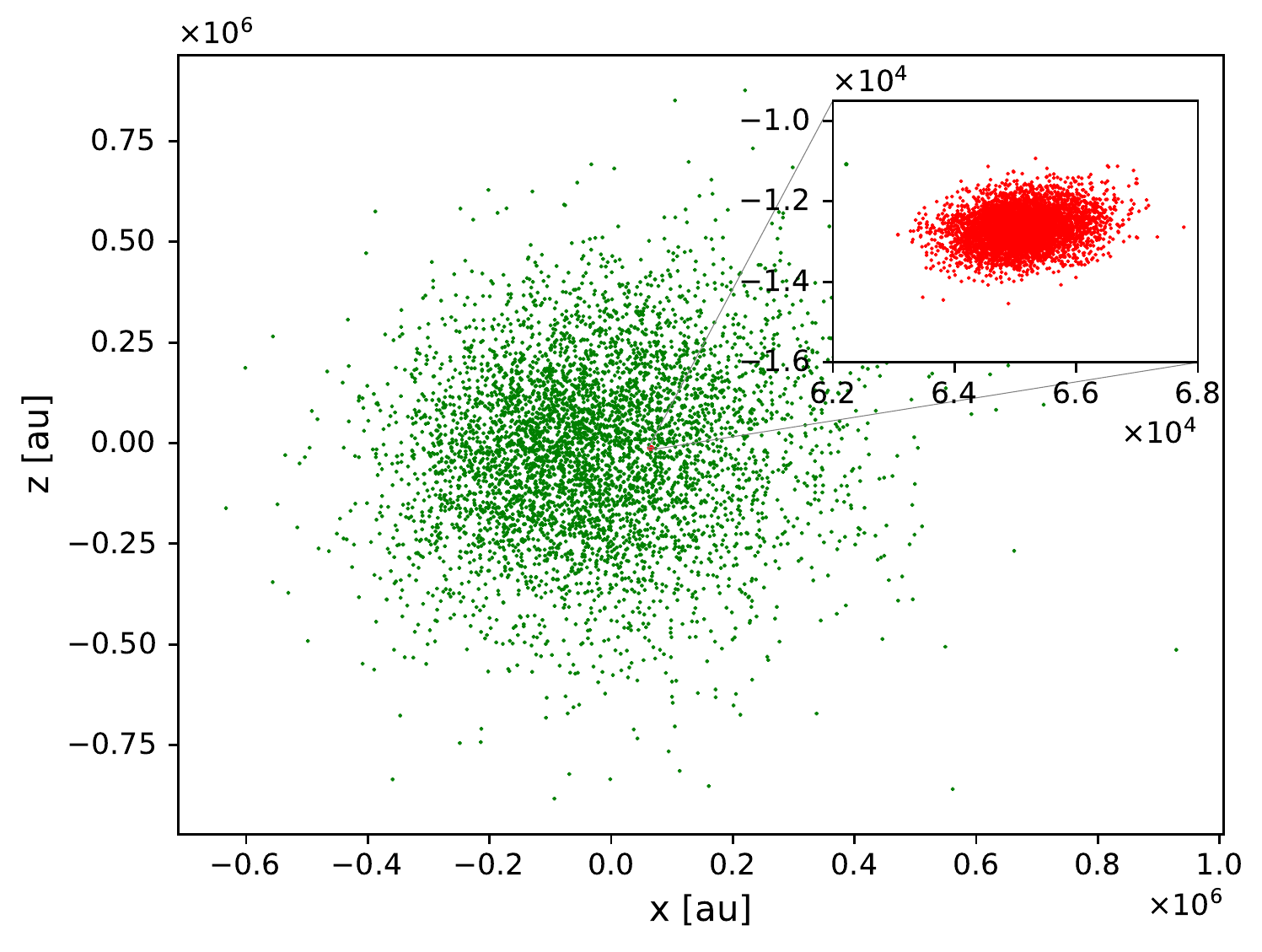}
    \caption{Spatial distribution of the clones of Star~B (green dots) and clones of HD~7977 (red dots) which were stopped during closest passage to nominal comets (C/2012\,F3 and C/2002\,A3 respectively) plotted in the Galactic heliocentric frame. Note the difference in scale of the two panels.}
    \label{fig:230_507_xz}
\end{figure}

In the absence of any stellar perturbations the comet C/2012~F3 can be traced back to its previous perihelion some 4.2\,Myr ago. During the observed apparition it has a perihelion distance of $q = 3.45$\,au. At the previous perihelion this value was even smaller: $q_{\rm{prev}} = 2.41$\,au when only Galactic tides are taken into account.

The situation changes drastically if we include the gravitational influence from Star B into the past motion model. As is presented in the right panel in Fig.~\ref{fig:2002a3_0230_2012f3_0507}, the orbital period of C/2012~F3 remains almost unchanged but the nominal previous perihelion distance now equals almost 7\,500\,au. And similarly to the case of C/2002~A3 described in Sect.~\ref{sec:2002a3}, the eccentricity appears to be much smaller before C/2012~F3 met Star~B. 

\subsection{The influence of the star on the swarm of cometary orbits}\label{sec:2012F3_star_VCs}

As in the previous example we traced numerically all 5001~VCs backwards under the influence of the Galactic potential and the passing Star B. In Fig.~\ref{fig:qe_2012f3_1Myr_2Myr}  we present two snapshots of the VCs' swarm $q$ -- $e$ distribution: 1\,Myr ago (left panel in the figure) and 2\,Myr ago (right panel). In this example the closest star--comet approach takes place at $-$1.64\,Myr. 

In contrast to the previous case the stellar path lies further from the comet's nominal orbit so now the outlying VCs (red points in the right panel of Fig.~\ref{fig:qe_2012f3_1Myr_2Myr}) are perturbed more than the nominal -- some of these outlying VCs passed much closer to the star path than the nominal, which can be seen also in the right panel of Fig.~\ref{fig:histograms}, where the distribution of minimal distances between the nominal star path and the comet clones is presented.

Even though C/2012~F3 has the highest possible orbit quality and therefore a very compact swarm of original VCs' orbits, the distribution shown in the right panel of Fig.~\ref{fig:histograms} shows still a sizable spread of minimal star -- VC distances and a significant part of the swarm has the minimal distance smaller than 500\,au. This, together with the geometry of the close approach, implies that practically we know nothing on this comet orbit before it met Star B. In fact 2 Myr ago an arbitrary perihelion distance was possible as well as an eccentricity from 0.4 to 1.0.

Again we have to stress that the orbital evolution reconstructed above {\it is traced backwards in time}. It means that in reality comet C/2012~F3 met Star B while moving in one of thousands of different orbits depicted as dots in the right panel of Fig.~\ref{fig:qe_2012f3_1Myr_2Myr} about 2\,Myr ago and then was observed in the orbit depicted with the respective dot from the left panel of this figure.

It is worth mentioning, that we have not detected any significant stellar perturbation for two the remaining orbital solutions of C/2012~F3 shown in Table~\ref{tab:2012f3_2002a3_models}.

\subsection{Star--comet encounter and stellar data uncertainties}\label{sec:2012F3_star_comet_full}

The situation is even worse when we study the influence of Star B uncertainties on our result. 

As for the previous pair of C/2002~A3 and HD\,7977 we decided to study this influence of the stellar data uncertainties for the C/2012\,F3 and Star B pair. After 5000 clones of the star were drawn it became obvious that we were dealing with a completely different case. For the discussed star errors of the proper motion are ten times bigger than the proper motion itself. Also the error of parallax is significantly bigger than in the case of HD\,7977. Because of the poor quality of the Star B astrometry published in Gaia DR2, an investigation of the swarm of clones of Star B acting on the nominal of C/2012\,F3 gives completely different results than those obtained in sect.~\ref{sec:2002a3}. While minimal distances of the cometary clones shown in Fig.~\ref{fig:histograms} are included in range from 400\,au up to 1\,800\,au, now the minimal distance between the star clones and the comet nominal vary from 550\,au to over 900\,000\,au. 

Only for the nominal Star B path the distance to any of the VCs is smaller than 1000\,au and only for another six (out of 5000) the minimal distance is less than 10\,000\,au. This of course means that there is a very small probability of such a close encounter in the past. But we should keep in mind that the nominal kinematic data are still the most probable one.

Contrary to the case discussed in sect.~\ref{sec:2002A3_star_comet_full} we can state that our knowledge on the motion of the star is much worse than that of the comet. 

To show how different are these two cases we prepared a plot with the comparison of the spatial distribution of stellar clones drawn from HD~7977 and Star B data. In Fig.~\ref{fig:230_507_xz} both these swarms of stellar clones are shown, all are stopped when their minimal distance from the nominal of the comet is reached.  The clones' positions are projected onto the $XZ$ plane of the Galactic heliocentric frame where the biggest scatter level is observed. It can be seen that the difference between the dispersion of both swarms is huge and this is due to the uncertain astrometry of Star B. This explains why the integration of clones of that star with the nominal of C/2012~F3 do not give similar results to those described in Sec.~\ref{sec:2002a3} and why the minimal distance between them can vary so much.

\section{New perspectives for studying LPCs' dynamics and origin}\label{sect:new_persp}

Our future work will focus on the search for other strong stellar perturbations which could influence cometary motion. The Gaia EDR3 and Gaia DR3 catalogues\footnote{see: https://www.cosmos.esa.int/web/gaia/release} are expected to improve astrometry and bring more astrophysical data including stellar mass estimates. This will extend the possibility to study stellar perturbations of cometary motion. 

The second mentioned catalogue is additionally supposed to include data on non-single stars which will fill an important gap in potential stellar perturbers data. All  searches for stars approaching the Sun published so far  are concentrated on single stars approach, even if it is already known that they are members of multiples. This almost always led to misleading results mainly because of proper motion contamination. While separately considered stars, which allegedly are parts of more complex systems, can give promising results (small minimal heliocentric distance), when we calculate the centre of mass of a system and use it instead of single stars it can change the result drastically, as the centre of mass can move in the other direction than the instantaneous movement of the stars it consists of. Therefore, to conduct a comprehensive research on close stellar approaches to the Sun, it is necessary to fully examine cases of binary and multiple systems. We hope that it will be possible thanks to new data included in upcoming catalogues.

We expect also a significant increase in radial velocity measurements, both in number and precision, necessary for our study. In the same time an improved quality of astrometry should guarantee better reliability of our calculations. 

\section{Conclusions}\label{sec:conclusions}
The main purpose of this paper was to present and examine the first detected real cases in which strong stellar perturbations significantly change the past motion of observed long period comets. Our first results show that such events are extremely rare. We have analyzed the influence of 647 potential stellar perturbers (i.e. all known to date) on 277 different LPCs and detected only two such cases.

In the presented examples a particular star acts as a significant perturber only on one particular comet since a great proximity is necessary. Additionally, these strong stellar perturbations occur only when a list of specific requirements is fulfilled. Significant stellar perturbations are possible only when a considered star has a small relative velocity at the time of the approach to the comet, preferably is massive, and the approach is very close, below 1000\,au. 

Apart from a detailed dependency on the geometry the overall strength of a stellar perturbation might be expressed as a velocity impulse size, see for example \cite{dyb-impuls:1994} for a detailed discussion. In short, an arbitrary orbit change will be possible when a velocity impulse gained by a comet from a star is comparable to a comet orbital velocity at the moment of the closest approach. It is easy to show, for example by means of a classical impulse approximation, that in both described cases the minimal star -- comet distance is small enough to induce such a significant velocity change. Instead we used a much more precise method, namely strict numerical integration, which allowed us to show the discussed cases in detail.

For a particular star approaching the Solar System all necessary conditions to significantly perturb any of the observed LPCs are fulfilled very rarely, in this investigation only twice for 647x277 analyzed cases. Moreover, the uncertainties of both stellar and cometary data make recognizing events that really happened in the past very difficult.

To obtain a reliable picture of the past comet dynamics we need reliable data. Our first example shows that for cases of high quality stellar data (the case of HD\,7977) our knowledge of a particular LPC dynamical history might be limited by its orbital uncertainties which cannot be improved after the last positional observation of that comet was obtained.

But our second example carries a different message. Although our study is based on the currently best available stellar data - Gaia DR2 it can be easily seen that not all of them have equal quality and can be successfully used for our purpose. The available astrometry of Gaia DR2 5700273723303646464 is of poor quality and do not clarify whether stellar perturbations could occur in this case. Due to astrometry uncertainties, especially large errors of proper motion, we could not properly test the reliability of the approach and therefore confirm the occurrence of strong stellar perturbations. 

Anyway we found our results to be the first confirmation of the Oort concept using real stars that make close approaches and significantly change the cometary orbits. All earlier works on that subject were based only on simulations \citep[see for example][]{fouchard_f_r_v:2011}.

We are determined to continue our research on close stellar passages and their influence on LPCs' dynamical history, still waiting for more numerous and better quality stellar data.

\section*{Acknowledgements}

We are very grateful to the reviewer, Luke Dones, for insightful comments that have helped us improve this paper.

We have made use of the SIMBAD database, operated at CDS, Strasbourg, France and also  of the VizieR catalogue access tool, CDS, Strasbourg, France (DOI: 10.26093/cds/vizier). The original description of the VizieR service was published in A\&AS 143, 23. This work has made use of data from the European Space Agency (ESA) mission {\it Gaia} (\url{https://www.cosmos.esa.int/gaia}), processed by the {\it Gaia} Data Processing and Analysis Consortium (DPAC, \url{https://www.cosmos.esa.int/web/gaia/dpac/consortium}). Funding for the DPAC has been provided by national institutions, in particular the institutions participating in the {\it Gaia} Multilateral Agreement.

This research was partially supported by the project 2015/17/B/ST9/01790 founded by the National Science Centre in Poland.

\bibliographystyle{mnras}

\bibliography{moja25}

\label{lastpage}

\end{document}